\documentclass{aa}

\usepackage{times}
\usepackage{graphicx}

\title{A layered model for non-thermal radio emission\\ from single O stars}
\author{S. Van Loo \and M.C. Runacres \and R. Blomme}
\institute{Royal Observatory of Belgium, Ringlaan 3, B-1180 Brussel, Belgium}
\offprints{S. Van Loo,\\ \email{Sven.VanLoo@oma.be}}

\date{Received  / Accepted}

\abstract{
We present a model for the non-thermal radio emission from bright O stars, in terms
of synchrotron emission from wind-embedded shocks. The model is an
extension of an earlier one, with an improved treatment of the cooling of
relativistic electrons. This improvement limits the synchrotron-emitting
volume to a series of fairly narrow layers behind the shocks. We show that the
width of these layers increases with increasing wavelength, which has
important consequences for the shape of the spectrum. We also show that the
strongest shocks produce the bulk of the emission, so that the emergent radio
flux can be adequately described as coming from a small number of shocks, or
even from a single shock.

A single shock
model is completely determined by four parameters: the position of the shock,
the compression ratio and velocity jump of the shock, and the surface
magnetic field.
Applying a single shock model to the O5 If star \object{Cyg OB2 No. 9} allows a
good determination of the compression ratio and shock position and, to a
lesser extent, the magnetic field and 
velocity jump.

Our main conclusion is that strong shocks need
to survive out to distances of a few hundred stellar radii. Even 
with multiple shocks, the shocks needed to explain the observed emission are
stronger than predictions from time-dependent hydrodynamical simulations.

}

\begin{document}
\maketitle

\section{Introduction}
About one quarter of the brightest O stars have a detectable radio flux of
non-thermal origin, in addition to their thermal radio emission (Bieging et al.
\cite{BAC89}).  The {\em thermal} radio emission is due to free-free emission 
in the stellar wind (Wright \& Barlow \cite{WB75}), whereas the {\em non-thermal} 
radiation is probably synchrotron emission by shock-accelerated electrons 
(White \cite{W85}). Usually, the thermal emission is only a small fraction of the
flux, and the non-thermal emission dominates at centimetre wavelengths.  

In a previous paper (Van Loo et al.~\cite{VL04}, hereafter Paper~I) we 
used a phenomenological model to explain 
the observed spectrum of the O5 If star \object{Cyg OB2 No. 9}.  The synchrotron-emitting 
electrons are assumed to be accelerated by wind-embedded
shocks, such as those produced by the instability of the radiative driving
mechanism. A power law was assumed for the momentum distribution of the
electrons and the synchrotron-emitting volume was assumed to extend to an outer
boundary $R_{\rm max}$. This approach leads to a model with a small number of
free parameters, all of which are well-determined by the fit to the
observations. Particularly, the interplay between non-thermal emission and
thermal absorption places a firm constraint on the outer boundary $R_{\rm max}$.

In Paper~I, the spatial distribution of the relativistic electrons was given
by a power law. This implies that the radiation is emitted by a continuous
volume extending out to $R_{\rm max}$. In the present paper, we take a closer
look at this assumption. In the shock-acceleration mechanism discussed in
Paper~I (first order Fermi acceleration, Bell \cite{B78}) electrons are 
accelerated to relativistic energies by making many round-trips over the shock.
Relativistic electrons are subject to a number of cooling mechanisms, most notably 
inverse Compton scattering in the presence of a dense radiation field 
(Chen \cite{C92}).
They may therefore lose their energy rather fast when they eventually escape 
downstream and move away from the shock. In this case the emission is not
expected to come from a continuous volume, but rather from a number
of (possibly thin) layers behind the shocks.

Hydrodynamical models (e.g. Runacres \& Owocki \cite{RO04}) of instability-generated
shocks show a wide variety of shock strengths (both in 
velocity jump
and
compression ratio). We find that the strongest shocks generate most of the
synchrotron emission, which means that the number of layers responsible for
the observed synchrotron emission can actually be quite small.

In the following, we first (Sect. \ref{sect:emissionlayers}) derive the momentum
distribution of relativistic electrons in the presence of cooling and
discuss the effect on the emergent flux. In Sect. \ref{sect:application}, we then 
apply a model where the emission is concentrated in a small number of thin layers to radio
observations of the same star which was used in Paper~I, \object{Cyg OB2 No. 9}.
Finally (Sect. \ref{sect:conclusions}),
we discuss the implications of our results and draw some conclusions.

\section{Emission layers}\label{sect:emissionlayers}

\subsection{Momentum distribution at the shock front}\label{sect:atshock} 
Electrons are accelerated at a shock by the first order Fermi acceleration
(Bell~\cite{B78}). Every time an electron crosses the shock front, it gains a
small amount of momentum. When an electron is kept in the vicinity of the
shock front and crosses the shock front many times, it can be accelerated
to relativistic momenta. The acceleration rate of the Fermi mechanism is given 
by
\begin{equation}\label{eq:accelerationrate}
        \frac{{\rm d}p}{{\rm d}t}=\frac{\Delta p}{\Delta t_{\rm cycle}},
\end{equation}
where
\begin{equation}
        \Delta p=\frac{4}{3}\frac{\Delta u}{c}p,
\end{equation}
is the momentum gain for an electron per round-trip. 
The round-trip time $\Delta t_{\rm cycle}$ can be determined as follows.
 After crossing the shock, a typical electron travels a mean free path 
$\lambda=pc/eB$ (White \cite{W85}), where $B$ is the magnetic field and other 
symbols have their usual meaning,  before it is scattered by magnetic 
irregularities. An electron does not necessarily return to the shock 
after one scattering. From Lagage \& Cesarsky~(\cite{LC83}) we derive that an 
electron is 
scattered $c/2u$ times
before recrossing the shock, where $u$ is 
the speed of the bulk motion upstream ($u_1$) or downstream ($u_2$) of the 
shock (both measured in the shock frame). Since the time between scattering events 
is given by $2\lambda/c$, the time needed for an electron to cross the shock is 
$\lambda/u$. Therefore the round-trip time is given by (Lagage \& Cesarsky \cite{LC83})
\begin{equation} \label{eq:roundtriptime}
        \Delta t_{\rm cycle}=\frac{\lambda}{u_1}+\frac{\lambda}{u_2}.
\end{equation}
For a shock with a compression ratio $\chi=u_1/u_2$ and a velocity jump 
$\Delta u=u_1-u_2$, the speeds (measured in the shock frame) are given by 
\begin{eqnarray}\label{eq:outflow}
	u_1&=&\frac{\chi\Delta u}{\chi-1},  \nonumber \\
	u_2&=&\frac{\Delta u}{\chi-1}. 
\end{eqnarray}
Then 
\begin{equation}\label{eq:roundtriptimeb}
        \Delta t_{\rm cycle}=\frac{pc}{eB}\frac{\chi^2-1}{\chi\Delta u}.
\end{equation}
The round-trip time $\Delta t_{\rm cycle}$ given in Eq.~(\ref{eq:roundtriptime}) 
is different from the expression $\Delta t_{\rm cycle}=2\lambda/c$ given by
Chen~(\cite{C92}). As this is the expression for the mean time between 
scatterings, the underlying assumption in Chen (\cite{C92}) is that the electron
returns to the shock after one scattering.

Electrons are accelerated to high momenta, but also lose their momentum via
different energy loss mechanisms. By taking the cooling into account, the net
momentum gain rate of the acceleration process is drastically reduced. At the
high momentum end of the spectrum, electrons are mainly cooled by the 
inverse-Compton scattering of photospheric UV photons at a rate 
(Rybicki~\&~Lightman~\cite{RL79})
\begin{equation}\label{eq:ICcooling}
        \left(\frac{{\rm d}p}{{\rm d}t}\right)_{\rm IC}=
	-\frac{\sigma_{\rm T}L_*}{3\pi r^2 c}
	\left(\frac{p}{m_{\rm e}c}\right)^2,
\end{equation}
where $\sigma_{\rm T}$ is the Thomson cross section, $L_*$ the stellar
luminosity and $r$ is the distance from the star.

The net momentum gain per time interval due to first order Fermi acceleration
and inverse-Compton cooling is then given by
\begin{equation}\label{eq:totalmomentumrate}
        \left(\frac{{\rm d}p}{{\rm d}t}\right)_{\rm net}=
        \left(\frac{{\rm d}p}{{\rm d}t}\right)_{\rm acc}
         +\left(\frac{{\rm d}p}{{\rm d}t}\right)_{\rm IC}.
\end{equation}
The highest momentum attainable for relativistic electrons is determined by the 
balance between acceleration and inverse-Compton cooling at the shock. The result is
a high momentum cut-off $p_c$ given by
\begin{equation}\label{eq:theoreticalpc}
        p_c=m_{\rm e}c\left[\frac{\chi \Delta u^2}{\chi^2-1}
        \frac{4\pi R_*^2}{\sigma_{\rm T}L_*} \frac{eB_*}{c}
	\frac{v_{\rm rot}}{v_\infty}\frac{r}{R_*}\right]^{1/2},
\end{equation}
where $R_*$ is the stellar radius, $v_{\rm rot}$ the rotational velocity of
the star and $v_\infty$ the terminal velocity of the stellar wind and
where we used $B=B_*\frac{v_{\rm rot}}{v_\infty}\frac{R_*}{r}$
(Weber~\&~Davis~\cite{WD67}). Using typical hot-star values 
($L_*=10^6~{\rm L_\odot}$, $v_\infty=2\,500~{\rm km\,s^{-1}}$, 
$v_{\rm rot}= 200~{\rm km\,s^{-1}}$, $R_*=20~{\rm R_\odot}$ and 
$\Delta u=100~{\rm km\,s^{-1}}$), we can estimate a numerical value for the 
high momentum cut-off,
\begin{equation}\label{eq:numericalpc}
        p_c\approx \frac{15~{\rm MeV}}{c}\sqrt{\frac{B_*\chi}{\chi^2-1}\frac{r}{R_*}}.
\end{equation}
Note that $p_c$ increases further out in the wind, due to the rapid
depletion of UV photons. The high momentum cut-off is also weakly dependent
on the compression ratio. Electrons can attain higher momenta in weak shocks,
because the acceleration rate is higher in these shocks, as can be seen from
Eqs.~(\ref{eq:accelerationrate}) and (\ref{eq:roundtriptimeb}). 
The high momentum cut-off is  a factor of $\sqrt{c/\Delta u}\approx 30$ 
lower than the value derived in Chen~(\cite{C92}) and in Paper~I, due to 
the different expressions for $\Delta t_{\rm cycle}$ (Eq. (\ref{eq:roundtriptimeb})).

In the absence of cooling, the acceleration of electrons through
first order Fermi acceleration results in a  power law momentum distribution
(Bell~\cite{B78}). When cooling is taken into account, the momentum 
distribution can be quite different from the power law. Near the high 
momentum cut-off, the net momentum gain goes to zero and the distribution will 
deviate from a pure power law. At low momenta, however, the acceleration rate is 
much higher than the cooling rate, so the momentum distribution is still a power 
law. Chen~(\cite{C92}) showed that the spectrum including Compton cooling can 
be expressed as
\begin{equation}\label{eq:distribution}
        N(p)=N_Ep^{-n}\left(1-\frac{p^2}{p_c^2}\right)^{(n-3)/2},
\end{equation}
where 
\begin{equation}\label{eq:nchi}
	n=\frac{\chi+2}{\chi-1},
\end{equation}
and $N_E$ is a normalisation constant related
to the energy density of the relativistic electrons (see below). 
The shape of the distribution is shown by the thick solid line in 
Fig. \ref{fig:behindshock}. The hook at the high momentum cut-off corresponds 
to the accumulation of electrons that, in the absence of cooling, would have 
had momenta above $p_c$. (A downward hook corresponding to the depletion of very 
energetic electrons can also occur. For a detailed explanation see 
Van Loo (\cite{Sven}).) As the shape of the momentum distribution does not depend on the round 
trip time, Eq. (\ref{eq:distribution}) is unchanged from Chen (\cite{C92}).

We determine the normalisation constant $N_E$ by specifying the 
amount of energy that goes into the relativistic electrons. Following 
Ellison \& Reynolds~(\cite{ER91}), we assume that the relativistic electrons
contribute little to the overall momentum balance across a shock. This means 
that the pressure of the relativistic electrons is only a small fraction 
($\zeta$) of the total pressure difference. It is unlikely that $\zeta$ is more 
than 0.05 and it could be conceivably less (Blandford \& Eichler~\cite{BE87}; 
Eichler \& Usov~\cite{EU93}). From the Rankine-Hugoniot relations we find that
the pressure difference across the shock is
\begin{equation}\label{eq:momentumbalance}
        \Delta P= \rho_1\frac{\chi}{\chi-1}(\Delta u)^2.
\end{equation}
Here $\rho_1$ is the density of the preshock gas. For simplicity we assume
$\rho_1=\dot{M}/4\pi r^2 v_\infty$. The relativistic electron pressure
is given by (White~\cite{W85})
\begin{equation}\label{eq:relpressure}
        P_{\rm rel}=\frac{1}{3}U_{\rm rel}=
	\frac{1}{3}\int_{p_0}^{p_c}{{\rm d}p\, p c N(p)},
\end{equation}
with $U_{\rm rel}$ the energy density of the relativistic electrons and 
$p_0$ the low momentum cut-off. For reasons that will become apparent in
Sect. \ref{sect:du}, we set $\zeta$ {\rm equal to} 0.05 (as opposed to 
$\zeta \leq 0.05$). Using Eqs.~(\ref{eq:momentumbalance}) and 
(\ref{eq:relpressure}), we can then determine $N_E$.

The normalisation of the momentum distribution differs from the one used
in Paper~I, which was based on the relativistic number 
density being smaller than the total electron number density.
For a given momentum distribution, the total number of relativistic electrons
and the total energy are of course directly related. A constraint on one 
implies a constraint on the other. The energy constraint is, in principle,
the more stringent as it is reached well before all electrons become 
relativistic.

\subsection{Momentum distribution behind the shock front} 
Once  an electron leaves the shock front, the acceleration ceases and 
cooling mechanisms will reduce its momentum. This means that the
momentum distribution of relativistic electrons behind the shock will differ
from the distribution observed at the shock. To describe the distribution
behind the shock, we need to know where the electrons were accelerated and
how their momentum is changed under the influence of cooling.

While the shock travels through the stellar wind with approximately the
terminal speed $v_\infty$, the relativistic electrons will move away from the shock 
front with the gas, at a significantly lower speed $u_2$ (see Eq. (\ref{eq:outflow})).
This means that outer wind electrons may well have been accelerated in the inner wind.
If the shock is at $R_{\rm S}$, the distribution of electrons at radius $r$, close to 
$R_{\rm S}$, was initially accelerated when the shock was at
\begin{equation}\label{eq:initialradius}
        r_{\rm i}= R_{\rm S}- v_\infty \frac{|R_{\rm S}-r|}{u_2}<R_{\rm S},
\end{equation}
where we have assumed that $v_\infty$ and $u_2$ are constant throughout the
wind. This expression is valid for both forward shocks (which move outward 
through the gas) and reverse shocks (which move inward through the gas). 
The momentum distribution produced at $r_{\rm i}$ is given by 
Eq.~(\ref{eq:distribution}) with $p_c$ and $N_E$ evaluated at $r_{\rm i}$.

Once we know where the electrons are accelerated, we only need to take into
account the cooling of the electrons. Inverse-Compton cooling, adiabatic 
cooling, synchrotron cooling and Bremsstrahlung cooling can all play a r\^ole. 
Since the observable radio emission comes from electrons with momenta of the 
order $\sim$ 100 MeV/c (see Paper~I), we will only incorporate the cooling
mechanisms which act at these momenta and higher. At high momenta, the 
electrons are mainly cooled by inverse-Compton cooling which dominates over
synchrotron cooling (Chen \cite{C92}). For electrons with intermediate 
momenta, the adiabatic cooling is more important than Bremsstrahlung 
(Chen \cite{C92}). We will therefore include only inverse-Compton and adiabatic
cooling. 

The cooling rate of the 
relativistic electrons is not only dependent on the momentum of the electrons, 
but also changes with distance. Since the relativistic electrons 
move along with the thermal gas, we will express the cooling rate as the loss 
of momentum over a distance ${\rm d}r$, where we take 
${\rm d}r=v_\infty {\rm d}t$. The cooling rate for relativistic electrons is then
given by (e.g. Chen~\cite{C92})
\begin{equation}\label{eq:totalcooling}
        \frac{{\rm d}p}{{\rm d}r}=-a_{\rm ic}\frac{p^2}{r^2}-\frac{2}{3}\frac{p}{r},
\end{equation}
where the first term represents the inverse-Compton cooling with 
$a_{\rm ic}=\sigma_{T}L_*/(3\pi m_{\rm e}^2 c^3 v_\infty)$  and the second 
term the adiabatic cooling. An electron which was initially at $r_{\rm i}$ with 
momentum $p_{\rm i}$ will be cooled down to a momentum (Chen~\cite{C92})
\begin{equation}\label{eq:coolingcurve}
        p=\frac{p_{\rm i}\left(\frac{r_{\rm i}}{r}\right)^{2/3}}
        {1+\frac{3a_{\rm ic}}{5}\frac{p_{\rm i}}{r_{\rm i}}\left[1-\left(
        \frac{r_{\rm i}}{r}\right)^{5/3}\right]},
\end{equation}
when it reaches $r$. 

The above expression enables us to determine the momentum 
distribution after cooling. 
We assume that a shock is spherically symmetric, i.e. that it covers a 
complete solid angle.
The number of relativistic electrons initially having 
a momentum between $p_{\rm i}$ and $p_{\rm i}+{\rm d}p_{\rm i}$ at a radius 
$r_{\rm i}$ is 
then
$4\pi r_{\rm i}^2 N(p_{\rm i}) {\rm d}p_{\rm i}{\rm d}r$, with 
$N(p)$ the differential number distribution given by Eq.~(\ref{eq:distribution})
and $4\pi r^2_{\rm i} {\rm d}r$ the volume that contains the electrons. 
Inverse-Compton and adiabatic losses reduce the momentum of the electrons to 
values between $p$ and $p+{\rm d}p$ where $p$ is given by 
Eq.~(\ref{eq:coolingcurve}). Also, the volume of the shell expands to 
$4\pi r^2 {\rm d}r$. Since the number of electrons in this momentum interval and 
volume is not changed, the number distribution after cooling $N_c(r,p)$ is given 
by 
\begin{equation}\label{eq:distributionbehind}
        N_c(r,p)=N(p_{\rm i}) \left(\frac{r_{\rm i}}{r}\right)^2
        \frac{{\rm d}p_{\rm i}}{{\rm d}p},
\end{equation}
where $p_{\rm i}$ is given by the inverse function of Eq.~(\ref{eq:coolingcurve}). 
Eq.~(\ref{eq:distributionbehind}) is valid for all momenta lower than the momentum 
$p_{c,e}$, to which an electron that had a momentum $p_c$ at $r_i$ has cooled at $r$.
Therefore, $p_{c,e}$ is given by Eq.~(\ref{eq:coolingcurve}) with $p_i=p_c$.
Note that the momentum distribution at the shock front, $N(p)$ in Eq.~(\ref{eq:distribution}), is given 
by $N_c(R_S,p)$. 

\begin{figure}
\resizebox{\hsize}{!}{\includegraphics{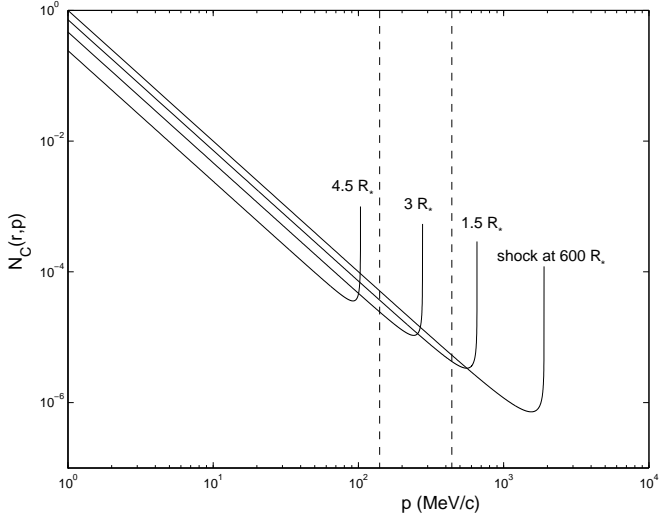}}
\caption{The cooling of relativistic electrons behind a strong shock ($\chi=4$
        and $\Delta u=100~{\rm km\,s^{-1}}$). The thick solid line is the momentum 
	distribution at the shock located at $600~R_*$. The three solid lines are the 
	distributions away from the shock, at locations $600~R_*-1.5~R_*$, $600~R_*-3~R_*$
	and $600~R_*-4.5~R_*$. We assumed $N(1~{\rm MeV/c})=1$ at $600~R_*$ and 
	$B_*=100~{\rm G}$. The dashed 
	lines indicate the momentum that the relativistic electrons must have to 
	produce synchrotron radiation between 2 and 20~cm 
	($450~{\rm MeV/c}>p>140~{\rm MeV/c}$).}
\label{fig:behindshock}
\end{figure}

An example for the momentum distribution is given in 
Fig.~\ref{fig:behindshock}. Here we plot the momentum distribution at the shock 
front for a strong shock ($\chi=4$ and $\Delta u= 100~{\rm km\,s^{-1}}$) at 
$600~R_*$ and the momentum distributions which are found $1.5~R_*$, $3.0~R_*$ 
and $4.5~R_*$ behind the shock. These distributions were accelerated respectively 
at $470~R_*$, $339~R_*$ and $209~R_*$. It is clear that the highest attainable 
momentum drops off rather rapidly behind the shock: electrons which are $4.5~R_*$
behind the shock no longer contribute to the observable radio emission, i.e.
the emission layer is no thicker than $4.5~R_*$. 

Eqs.~(\ref{eq:coolingcurve}) and (\ref{eq:distributionbehind}) are valid
for both reverse and forward shocks. 
The momentum distribution at a distance $|R_S-r|$ behind a reverse
or forward shock is initially accelerated at $r_i$. Electrons with a momentum $p_i$ at 
$r_i$ are then cooled down to a momentum $p$ at $r$. Although $r$ is not the same for a 
reverse and forward shock, the momentum $p$ is not significantly different as long 
as $|R_S-r|\ll|R_S-r_i|$. Since this criterion is fulfilled for $u_2\ll v_\infty$, 
this means that a reverse and forward shock will have the same momentum distribution 
at a distance $|R_S-r|$ downstream of the shock. It is therefore possible 
to discuss only forward shocks in the rest of the paper without loss of generality.

\subsection{Synchrotron emission layers}\label{sect:synemissionlayer}
The previous section shows that the relativistic electrons (and thus also the
synchrotron emission) are limited to narrow layers behind
the shock. The width of the synchrotron emission layers is determined by the distance 
over which the relativistic electrons still produce observable radio emission. The
emission at a given frequency produced by a distribution of relativistic electrons
is given by (Paper~I)
\begin{equation}\label{eq:emission}
        j_\nu(r)=\frac{1}{4\pi}\int_{p_0}^{p_{\rm c,e}}{{\rm d}p\, N_c(r,p) 
	\bar{P}_\nu(r,p)},
\end{equation}
where $\bar{P}_\nu (p)$ is the synchrotron power of a single electron averaged
over solid angle. We take into account the Razin effect (Razin~\cite{R60};
see also Ginzburg~\&~Syrovatskii~\cite{GS65}), which is a plasma effect
that greatly reduces the synchrotron emitting power.

Although the emissivity is calculated by integrating over all momenta, only
electrons within a narrow range of momenta contribute to the emissivity at
a given frequency. A relativistic electron with momentum $p$ emits most of
its energy in a narrow region around the frequency
(Ginzburg~\&~Syrovatskii~\cite{GS65})
\begin{equation}\label{eq:maxfrequency}
        \nu\approx 0.29 \frac{3}{4\pi}\frac{eB}{m_{\rm e}c}
	\left(\frac{p}{m_{\rm e}c}\right)^2.
\end{equation}
In a local magnetic field of $B=0.015~{\rm G}$ the synchrotron emission between 
2 -- 20 cm (15 -- 1.4 GHz) is produced by electrons with momenta between 140 and 
450~MeV/c. The emission  at a given wavelength is thus related to a specific momentum 
and the emission layer extends up to the distance behind the shock where all 
relativistic electrons are below this momentum.

The width of the synchrotron emission layer $\Delta R_S$ is 
determined by the time $\Delta t_{\rm cool}$ to cool the relativistic electrons 
below radio emitting momenta.  From Eq.~(\ref{eq:totalcooling}) we see that 
the cooling rate decreases with distance from the star. This means that the 
relativistic electrons are cooled faster close to the star and that the width 
of the emission layer behind the shock is larger further out in the wind. The width 
is proportional to $\Delta t_{\rm cool}$.  The cooling time $\Delta t_{\rm cool}$
in principle also depends on the value of the high momentum cut-off $p_c$, but this 
turns out to be a minor effect. Furthermore, the width of the emission layer is 
influenced by the outflow speed of the relativistic electrons. Because the 
relativistic electrons flow away from the shock with a velocity $u_2$ given
in Eq.~(\ref{eq:outflow}), the width of the emission layer is directly proportional
to $u_2$.

\begin{figure}
\resizebox{\hsize}{!}{\includegraphics{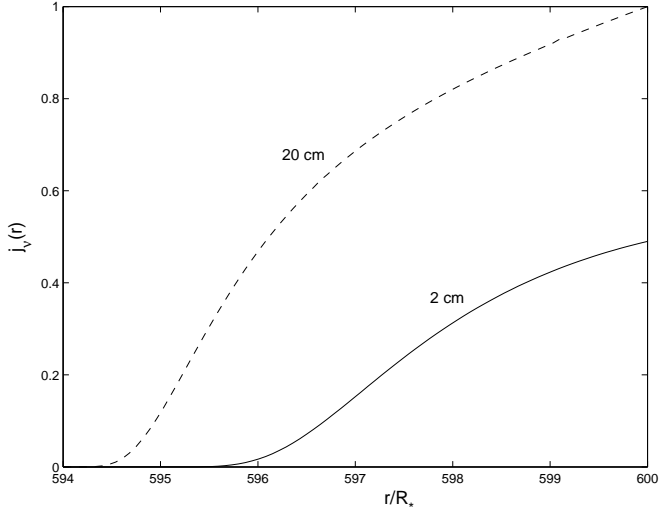}}
\caption{The emissivity behind a strong forward shock at $r=600~R_*$ for 2
        different wavelengths. The dashed line is the emissivity at 20~cm
        and the solid line at 2~cm. We normalised the emissivity to the
        highest value.}
\label{fig:emissionlayer}
\end{figure}

An important point is that the width of the emission layers depends on 
wavelength. At short wavelengths the radiation is produced by more energetic electrons 
(Eq.~(\ref{eq:maxfrequency})). Since these  electrons are cooled more rapidly by 
inverse-Compton cooling 
(Fig.~\ref{fig:behindshock}),  the emission layers at shorter 
wavelengths are narrower, as can be seen in Fig.~\ref{fig:emissionlayer}.
These differences change the shape of the synchrotron spectrum
(Sect.~\ref{sect:spectralindex}). 
The emission layers in Fig.~\ref{fig:emissionlayer} are somewhat larger than
expected from Fig.~\ref{fig:behindshock}, because the emission at a given wavelength
is produced by electrons, not with one specific momentum, but within a narrow 
momentum range.

\subsection{Emergent flux}\label{sect:emergentflux}
The emergent synchrotron flux of an electron distribution behind a shock 
is given by
\begin{equation}\label{eq:flux}
        F_\nu=\frac{1}{D^2}\int_{R_S-\Delta R_S(\nu)}^{R_S}{{\rm d}r 4\pi r^2 j_\nu(r)
                G\left(\frac{r}{R_\nu}\right)},
\end{equation}
where $D$ is the distance to the star, $R_\nu$ the characteristic radius of
emission (Wright~\&~Barlow~\cite{WB75}), $\Delta R_S(\nu)$ the width of the
emission layer at a frequency $\nu$ and $G(r/R_\nu)$ a function that describes the
influence of free-free absorption in the stellar wind (see Paper~I). Because 
of the triangular shape of the synchrotron emissivity behind a shock (see 
Fig.~\ref{fig:emissionlayer}), the synchrotron flux can be approximated, to first 
order, by
\begin{equation}\label{eq:approxflux}
        F_\nu\approx\frac{4\pi R^2_S}{D^2}
         \frac{j_\nu(R_S)G\left(\frac{R_S}{R_\nu}\right)}{2} \Delta R_S(\nu).
\end{equation}
All calculations are done using the exact expression 
Eq.~(\ref{eq:flux}), but this  approximate expression is useful in the discussions
of the following sections.

The synchrotron radio fluxes emitted behind a shock given in Eq.~(\ref{eq:flux})
can be calculated numerically.  In the code (which is similar to the one used 
in Paper~I), the momentum integral of the emissivity, given in Eq.~(\ref{eq:emission}), 
is computed using the trapezoidal rule and the step size is refined until it reaches 
a desired fractional precision of $10^{-3}$. The emissivity has to be calculated 
at different positions behind the shock. Then, the spatial integral of the 
flux, Eq.~(\ref{eq:flux}), is treated in the same way as the momentum integral, but 
up to a precision of order 1\%. Here the function $G(r/R_\nu)$ 
is determined using a linear interpolation between precalculated values.
The number distribution is given by Eq.~(\ref{eq:distributionbehind}), while 
the mean synchrotron power of a single electron $\bar{P}_\nu$ is determined
using precalculated values in the same way as $G(r/R_\nu)$.  We included 
the Razin effect in the calculation of $\bar{P}_\nu$ (see Paper~I). The low momentum cut-off 
is assumed $p_0=1~{\rm MeV/c}$, while the high momentum cut-off $p_{c,e}$ is 
found by evaluating Eq.~(\ref{eq:theoreticalpc}) in Eq.~(\ref{eq:coolingcurve}). 

The total emergent flux also has a contribution from free-free emission by 
{\it thermal} electrons. This contribution is calculated using the 
Wright \& Barlow formalism (Wright \& Barlow~\cite{WB75}) and added to
the non-thermal flux.

\subsection{Dominance of the strong shocks}\label{sect:dominance}
Hydrodynamical simulations show that a hot-star wind is filled with shocks
out to very large distances (Runacres~\&~Owocki~\cite{RO04}). These shocks
exhibit a large variation in
shock velocity jump and compression ratio.
In this section we will show that the shocks with high $\chi$ and $\Delta u$ dominate 
the radio emission. Consequently, the shock distribution responsible for 
the observed emission corresponds to a few isolated shocks in the wind.

The normalisation constant $N_E$ of the momentum distribution is 
proportional to the pressure difference $\Delta P$ across the shock, which 
has a $\Delta u^2$-dependence. This means that the number of electrons in all
momentum ranges increases with the 
shock velocity jump.
Furthermore the synchrotron 
emissivity $j_\nu$, given by Eq.~(\ref{eq:emission}), is directly proportional 
to $N_E$ which means that the emissivity must be proportional to $\Delta u^2$. In 
Sect.~\ref{sect:synemissionlayer} we showed that the width of the emission layer 
is also proportional to $\Delta u$. From Eq.~(\ref{eq:approxflux}) it then follows that
the synchrotron flux is proportional to $\Delta u^3$. This relation remains 
valid if we use the exact expression Eq.~(\ref{eq:flux}) to calculate the 
synchrotron flux. 

The radio fluxes do not only increase with the 
shock velocity jump,
but also with the compression ratio $\chi$. The slope of
the momentum distribution becomes shallower with higher $\chi$ (see Eq.
(\ref{eq:distribution})). Assuming that the 
number of relativistic electrons is the same near $p_0$ (i.e. $N_E$ independent
of $\chi$), more electrons are then present in the momentum range important 
for observable radio emission.
As the emissivity at a given wavelength is related to the number 
of electrons with a specific momentum, more emission is radiated by the 
strongest shocks. A complication arises because the normalisation constant 
$N_E$ actually decreases with $\chi$ (Sect. \ref{sect:atshock}), thereby slightly counteracting
the effect of the slope. However, this complication turns out to have a minor effect.
 A grid of models (discussed in Sect. \ref{sect:model}) shows that the flux only fails to increase
with increasing $\chi$ if the shock is close to the stellar surface and if at the same
time the magnetic field is strong. Even in such cases the effect is limited 
to 20 \% on the flux.

Thus, the shocks with the highest compression ratios and 
shock velocity jumps
in the wind dominate the radio emission.

\subsection{Interpretation of the spectral index}\label{sect:spectralindex}
The broadening of the synchrotron emission layers toward longer wavelengths
has consequences on the shape of the observed synchrotron flux: the
spectral index\footnote{The spectral index is defined as
$F_\nu\propto \nu^\alpha\propto\lambda^{-\alpha}$.} $\alpha$ of the emission is
steeper than expected. For the sake of clarity, we neglect the Razin
effect and the free-free absorption in the discussion.

For a power-law momentum distribution of relativistic electrons with index
$n$, the synchrotron emissivity $j_\nu$ is given by a power law in frequency with an 
exponent (Rybicki~\&~Lightman~\cite{RL79})
\begin{equation}\label{eq:exponent}
	\tilde{\alpha}=\frac{1-n}{2}.
\end{equation} 
Eq.~(\ref{eq:approxflux}), with $G(R_S/R_\nu)=1$, shows that the spectral index $\alpha$ 
of the synchrotron flux would equal $\tilde{\alpha}$ if the width of 
the emission layer were independent of frequency. 
But as the emission layers broaden toward longer wavelengths (see 
Sect.~\ref{sect:synemissionlayer}), the synchrotron flux increases at these 
wavelengths and the radio spectrum steepens. As a consequence the spectral 
index is lower than $(1-n)/2$.

In astrophysical applications the measured radio spectral index $\alpha$ 
is often used to estimate the compression ratio of the shock that accelerated 
the relativistic electrons (e.g. Chen~\&~White~\cite{CW94}). If we observe 
synchrotron emission with a radio spectral index $\alpha=-0.75$ then 
for a non-cooling model we would derive a compression ratio $\chi=3$ using 
$\alpha=\tilde{\alpha}$ and Eq. (\ref{eq:nchi}). However, taking into account 
cooling behind the shock, we find that the observed flux corresponds to
a strong shock with $\chi=4$.

\section{Application and results}\label{sect:application}

\subsection{Cyg OB2 No. 9: 21 Dec 1984 data}
\begin{table}[t]
 \caption{Relevant stellar parameters for \object{Cyg~OB2~No.~9} adopted in
        this paper. The numbers are taken from Bieging~et~al.~(\cite{BAC89}, BAC),
        Leitherer~(\cite{L88}, L) and Herrero~et~al.~(\cite{HCVM99}, HCVM).}
\label{tab:CygOB2No9}
\begin{center}
\begin{tabular}{lrl}
        \hline
        \hline
        Parameter & &Ref.\\

        \hline
        $v_{\rm rot} \sin i~({\rm km~s^{-1}})^a$&145&BAC\\
        $T_{\rm eff}~({\rm K})^b$&44\ 500&HCVM\\
        $R_*~({\rm R_{\sun}})$& 22&HCVM\\
        $d~({\rm kpc})$&1.82&BAC\\
        $v_\infty~({\rm km~s^{-1}})$&2\ 900&    L\\
        $\dot{M}~({\rm M_{\sun}~yr^{-1}})$&$2.0\times 10^{-5}$&L\\
        $L_*~({\rm L_{\sun}})$&$1.74\times 10^6$&HCVM\\
        \hline
\end{tabular}
\end{center}

$^a$ We follow White~(\cite{W85}) in assuming that
$v_{\rm rot}= 250~{\rm km~s^{-1}}$, which is not in contradiction with the value
for $v_{\rm rot} \sin i$. \\
$^b$ The wind temperature is assumed to be $0.3$ times the stellar effective
temperature (Drew~\cite{D89}). \\
\end{table}

In Paper~I we applied the continuous model to the non-thermal radio emitter
\object{Cyg~OB2~No.~9} (O5 If). Of all the observational radio data that are available
for \object{Cyg~OB2~No.~9}, we used the 1984 Dec 21 observation with the VLA
(Bieging et al.~\cite{BAC89}) because of its detection of emission in three
radio wavelength bands and the small error bars on the detection. However
the error bars on the \object{Cyg OB2 No. 9} observations listed by Bieging et
al.~(\cite{BAC89}) are surprisingly small (0.1 mJy). The error due to the
flux calibration should already be 5 \% of the observed flux for 2 cm
observations and 2 \% for 6 and 20 cm observations (Perley \&
Taylor~\cite{Perley+Taylor03}), which translates to 0.29, 0.15 and
0.10 mJy at 2, 6 and 20 cm. We can therefore conclude that only the
random errors (root-mean-square) were listed. If we include the 
calibration errors (absolute flux scale errors), we find:
5.7 $\pm$ 0.3 mJy (2 cm), 7.4 $\pm$ 0.2 mJy (6 cm) and 4.9 $\pm$ 0.14 mJy (20 cm).

\begin{figure} 
\resizebox{\hsize}{!}{\includegraphics{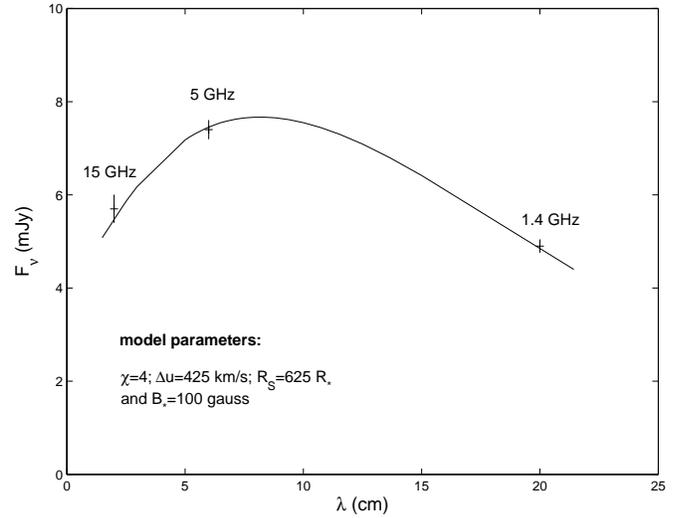}}
\caption{The VLA observations (21 Dec 1984) for \object{Cyg~OB2~No.~9} with revised 
	error bars. The solid line represents a single shock model that 
	explains the non-thermal radio emission. The stellar parameters are given in 
	Table~\ref{tab:CygOB2No9}.}
\label{fig:observation}
\end{figure} 

A significant fraction of the radio emission is not radiated as synchrotron
radiation by relativistic electrons, but is due to free-free emission from
the stellar wind (Wright~\&~Barlow~\cite{WB75}). Using the values from Table
\ref{tab:CygOB2No9}, the contributions
of the free-free emission\footnote{Contrary to Paper~I where we used $\mu=1.4$, 
$\gamma=1$ and $Z^2=1$, we use $\mu=1.3$, $\gamma=1.1$ and $Z^2=1.3$ in the Wright \& Barlow
formula.} to the observations of \object{Cyg~OB2~No.~9} are: 1.4 mJy
(2 cm), 0.7 mJy (6 cm) and 0.4 mJy (20 cm), if we assume that the stellar wind
is completely ionised. The stellar parameters adopted are listed in
Table~\ref{tab:CygOB2No9}. We will now apply a layered shock model to these
observations of \object{Cyg~OB2~No.~9}.

\subsection{Single shock model}\label{sect:model}
The bulk of the emission by a set of shocks is produced by its strongest members
(Sect.~\ref{sect:dominance}). To simplify the model further, we will consider the 
synchrotron emission from a single shock. The emission from multiple shocks can then 
be estimated by adding up the emission from single shock models (see 
Sect.~\ref{sect:multipleshocks}).  A single shock model is described by 
four parameters: the position of the shock $R_S$, the shock parameters 
$\Delta u$ and $\chi$ which determine the momentum distribution of the 
accelerated electrons and the surface magnetic field $B_*$. Hot stars are expected 
to have a surface magnetic field in the range of 10 to 200~G. The upper limit
corresponds roughly to the detection limit of the field (no Zeeman splitting has
been observed from any spectral line of any normal O star). The lower limit 
is set by the Razin effect, because otherwise no synchrotron radiation would be observed.

\begin{figure*}
\centering
\includegraphics[width=8.0cm]{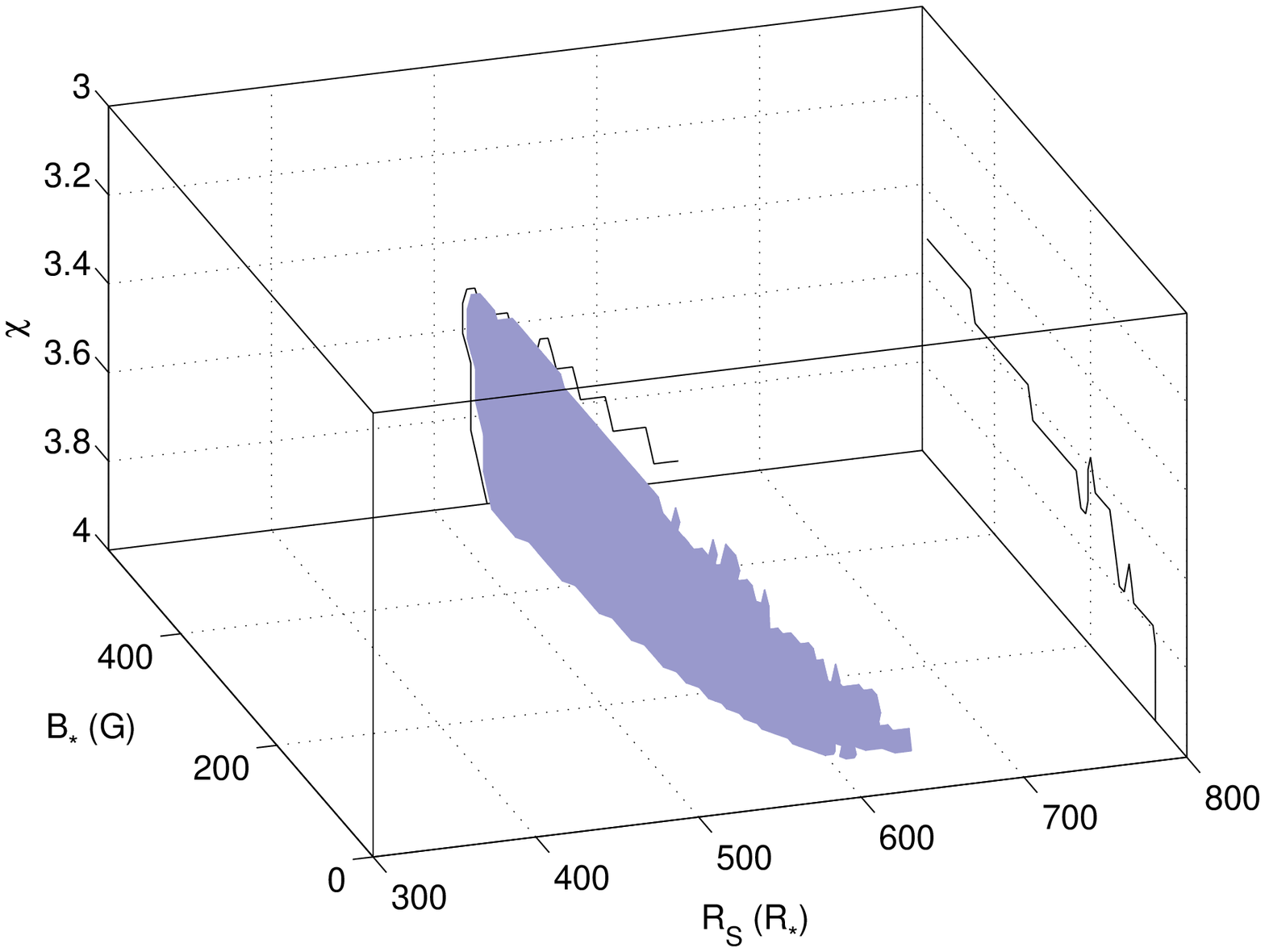}
\includegraphics[width=8.0cm]{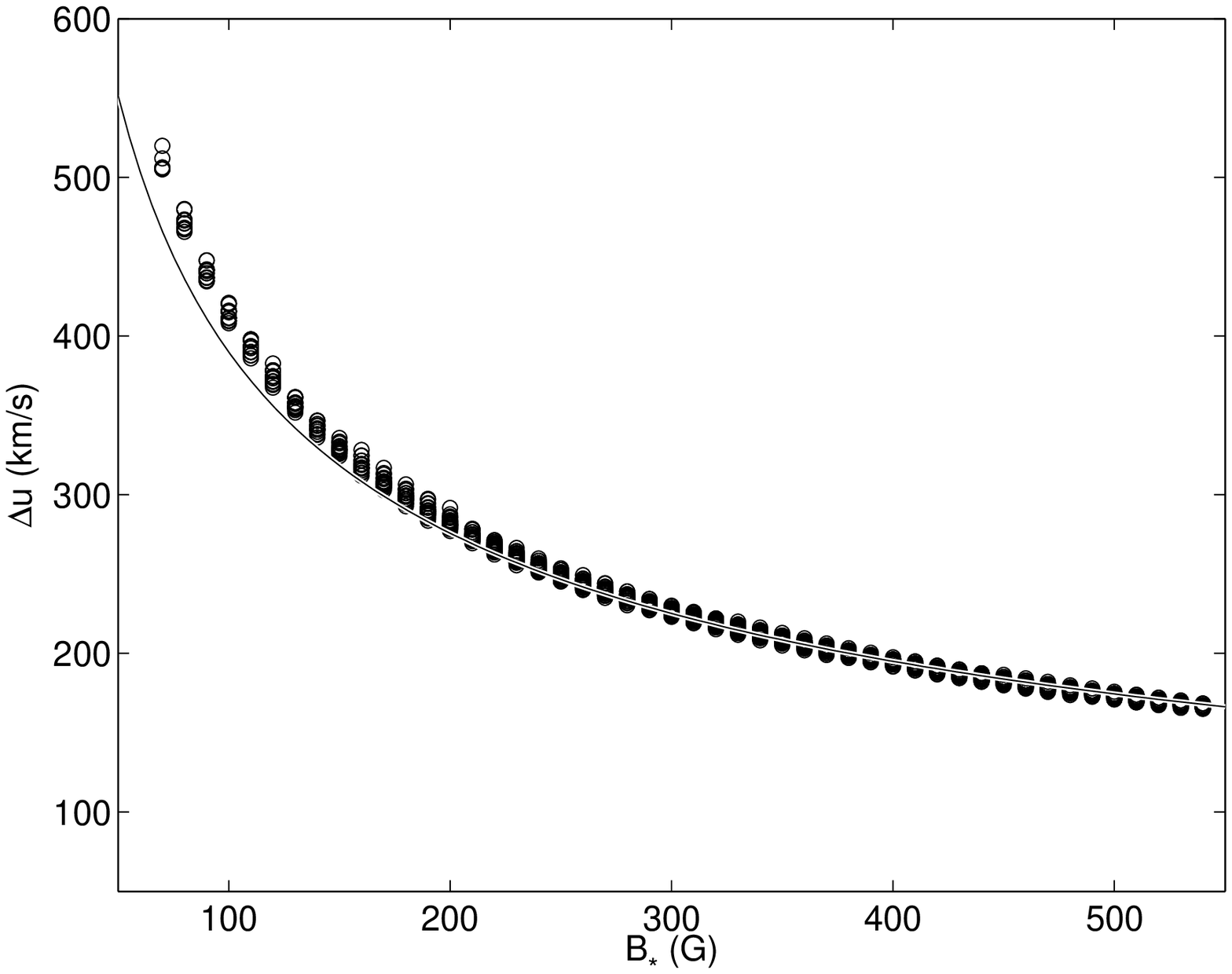}
  \caption{(a) {\em Left panel}: The combinations of $\chi$, $R_{\rm S}$ and $B_*$ that fit the
        observations for \object{Cyg~OB2~No.~9} lie in the 
	grey
	area on the
        figure. Projections on two planes are plotted to situate the solutions
        in the parameter space. (b) {\em Right panel}: For all possible combinations of 
	the model parameters that explain the
        radio observations of \object{Cyg~OB2~No.~9}, the values for the surface magnetic
        field $B_*$ and the 
	shock velocity jump
	$\Delta u$ are shown. The solid
        line represents $\Delta u^3  B_*^{3/2}$  is constant.}
  \label{fig:solutions}
\end{figure*}

Once a choice is made for the model parameters, we can calculate the 
synchrotron flux in the three wavelengths for a grid in $R_{\rm S}$, $\chi$, 
$\Delta u$ and $B_*$ and select those combinations of the parameters that 
fit the VLA observations within the error bars. 
The emergent flux for one such model is shown in Fig.~\ref{fig:observation}. 
All possible combinations of $R_{\rm S}$, $\chi$, $\Delta u$ 
and $B_*$ are shown in Figs.~\ref{fig:solutions}a and \ref{fig:solutions}b.
Strong constraints are produced on all four parameters. In principle 
the ranges of the model parameters depend on $p_0$, but we checked that 
the effect is minimal.

\subsection{Solution space}

\subsubsection{Compression ratio $\chi$ and shock position $R_S$}
Fig.~\ref{fig:solutions}a shows that the shock parameters $R_{\rm S}$ and $\chi$
are rather precisely determined. 
If the observations are to be explained by a single shock, then it must be
strong ($\chi=3.5-4$) and  located between $520-640~R_*$.

The reason that we find strong shocks is as follows. The emission at 
2 and 6 cm is barely influenced by the free-free absorption or
the Razin effect (see below). In a simple analysis, this means that, 
at these wavelengths, the exponent $\tilde{\alpha}$ of the synchrotron emissivity 
is simply the radio spectral index $\alpha$. We can then derive the compression ratio 
of the shock directly from the radio spectrum. Using
Eqs. (\ref{eq:nchi}) and (\ref{eq:exponent}), we then find $\chi=1-3/2\alpha$. 
(Note that this means that weaker shocks produce a steeper spectrum.) However,
the steepening due to the wavelength dependence of the emission layer (see Sect 
\ref{sect:spectralindex}) is not included in the above expression. This means that
the compression ratio derived for a layered model will be larger than the 
compression ratio derived in the simple analysis above.

The shock responsible for the observed emission at 20~cm must be located at a 
large distance from the star, due to the large
free-free absorption in the wind. All radio emission emitted too close 
to the star is absorbed and cannot be observed. This absorbing volume extends 
up to the radius $R_\tau(\nu)$ where the radial free-free optical depth is unity.
Using $\int_{R_\tau(\nu)}^{\infty}{\chi_{\rm ff} {\rm d}r}=1$ with 
$\chi_{\rm ff}$ the free-free absorption coefficient given by Allen (\cite{A73}), 
we find 
\begin{equation}
	R_\tau(\nu)=1.76\times10^{28} \left(\gamma g_\nu Z^2\right)^{1/3} T^{-1/2}
	\left(\frac{\dot{M}}{\mu v_\infty \nu}\right)^{2/3} {\rm cm},
\end{equation}
where $T$ is in K, $\dot{M}$ in $M_{\sun}$/yr, $\nu$ in Hz and $v_\infty$ in 
${\rm km\,s^{-1}}$ and where the other symbols have their usual meaning
(Wright \& Barlow \cite{WB75}).
Thus, the synchrotron emission must be produced beyond $R_\tau({\rm 20~cm})$ which is 
$450~R_*$. Radiation emitted beyond this distance is not 
absorbed at 2 and 6~cm because the free-free opacity is smaller at shorter 
wavelengths. Although a shock is needed at large distances, it cannot be too far out in 
the wind.  The free-free absorption at 20 cm would then be negligible, 
resulting  in a pure power-law spectrum that  does not explain the observations 
in Fig. \ref{fig:observation}.

The tight constraint on $R_{\rm S}$ presented here is only valid 
for a single shock. In Sect.~\ref{sect:multipleshocks}, we will explain the 
observations by multiple shocks. These can be spread out over a fairly large 
geometric region, and the shock positions are then much 
less well determined than in the case of a single shock.

\subsubsection{Surface magnetic field $B_*$}
The surface magnetic field $B_*$ is less constrained than the compression 
ratio and the location of the shock. We even find fits 
for the surface magnetic fields higher than 500 G, which is well above the
detection limit ($\approx 200~{\rm G}$). The reason for this is the influence of the 
magnetic field on the synchrotron spectrum through the Razin effect. The 
Razin effect is small for frequencies
\begin{equation}\label{eq:razineffect}
        \nu\gg 20\frac{n_e}{B},
\end{equation}
where $n_e$ is the number density of the thermal electrons (Ginzburg \&
Syrovatskii~\cite{GS65}). 
For a strong magnetic field the Razin effect is
negligible in the observed radio wavelengths, so that the synchrotron
spectral shape is only determined by the other two parameters, $R_{\rm S}$ and $\chi$.
For weaker magnetic fields, the Razin effect has some impact on the long wavelengths.
Since more radiation is already taken away by the Razin effect, 
less needs to be taken away by free-free absorption. This means that the shock must 
be somewhat further in the wind to explain the synchrotron spectrum. This can 
be seen from the bending of acceptable solutions for $B_*<200$~G in 
Fig.~\ref{fig:solutions}a.

\subsubsection{Shock velocity jump $\Delta u$}\label{sect:du}
Fig.~\ref{fig:solutions}b  shows the  relation between the
surface magnetic field $B_*$ and the 
shock velocity jump
$\Delta u$. The 
 shock velocity jump
increases from $170~{\rm km\,s^{-1}}$ (for $B_*=500$~G) to $500~{\rm km\,s^{-1}}$
(for $B_*=70$~G). Recall that we assumed $\zeta=0.05$ (Sect.~\ref{sect:atshock}).
For $\zeta<0.05$, the 
shock velocity jump
will be higher.

The reason for the tight relation is the direct dependence of the emission on 
both $\Delta u$ and $B_*$. In the absence of the Razin effect, the 
synchrotron emissivity
produced by a power-law momentum distribution is proportional to 
$N_E B_*^{(n+1)/2}$ (Rybicki \& Lightman~\cite{RL79}). In a lower magnetic field,
the relativistic electrons emit less synchrotron radiation and the number of 
electrons must increase (through $N_E$) to produce the same emission. 
Because $\chi$ and $R_S$ are nearly constant for all solutions (see  
Fig.~\ref{fig:solutions}a), $N_E$ can only be increased by increasing  
the 
shock velocity jump
$\Delta u$. All our solutions have 
$n=(\chi+2)/(\chi-1)\approx 2$, so to first order the synchrotron 
flux $F_\nu\propto\Delta u^3 B_*^{3/2}$ (see Fig.~\ref{fig:solutions}b). This 
largely explains the tight relation. The small difference for $B_*<200$~G 
is due to the Razin effect.

\subsection{Multiple shocks}\label{sect:multipleshocks}
The results we find for this single shock model can be re-interpreted
in terms of multiple shocks.
As in the case of a single shock, we assume that each shock covers a complete
solid angle.
 The emission of the multiple shocks is 
calculated by adding up the emission of single shock models.  
From Sect.~\ref{sect:dominance}, we know that the flux is proportional to $\Delta u^3$.
Multiple shocks, each with a lower $\Delta u$, can therefore also fit the fluxes provided
the sum of their $\Delta u^3$ equals the single shock $\Delta u^3$.
As a quantitative example we calculate a multiple shock model with 
a surface magnetic field near the detection limit ($B_*=200~{\rm G}$). The 
shock velocity jump
required for a single $\chi=4$ shock is then $290~{\rm km\,s^{-1}}$ (see 
Fig. \ref{fig:solutions}b).  As a reasonable value for wind-embedded shocks 
is $\Delta u=50~{\rm km\,s^{-1}}$ (e.g. Runacres \& Owocki \cite{RO04}), this means 
that we need about 200 shocks to produce the observed flux levels. 
A numerical calculation shows that 
a multiple shock model with 219 shocks (with $\chi=3.7$ and $\Delta u=50~{\rm  km\,s^{-1}}$) 
located between $50~R_*$ and $925~R_*$ indeed produces the observed spectrum.

The reason why we needed $\chi=3.7$ for the multiple shock model (as opposed to 
$\chi=4$ for the single shock model) is that free-free absorption partly counteracts the
steepening effect of the emission layer. This is because
the free-free absorption at 2 and 6 cm is no longer 
negligible for emission produced close to the star. For example, a shock near $100~R_*$
only contributes to the 2 cm flux and not to the 6 and 20 cm flux. A consequence is that 
the resulting spectrum of a multiple shock model is flatter than for a single shock model.  
This means that, to produce the observed spectrum, multiple shocks must be weaker than
derived in the single shock model. 

The above analysis shows that we can bring down the required velocity jump
by introducing a {\em large} number (in this case 219) of shocks. 
This assumes that all shocks have equal strength.
If the shocks have a range of strengths, the strongest shocks will dominate
(Sect.~\ref{sect:dominance}), and the emergent flux will be produced only by 
relatively {\em small} number of strong shocks.
This would make it more difficult to bring the typical
velocity jump down to a value consistent with hydrodynamical results. 
Whether the observed spectra can be reproduced by a model based on 
hydrodynamical predictions will be investigated in a subsequent paper 
(Van Loo et al. \cite{VL05}).

For magnetic fields lower than $200~G$, the shock velocity jump
derived in the single shock 
model is higher, hereby increasing the number of shocks needed to produce the observed 
fluxes. For $B_*=70~{\rm G}$ the number of shocks (with $\Delta u=50~{\rm km\,s^{-1}}$) needed 
is about 1000.  
Obviously such a high number is irreconcilable with hydrodynamical models.

\subsection{Comparison with continuous model}
In Paper~I the distribution of the relativistic electrons was described by
a power law both in momentum and distance. The normalisation of the 
distribution was done by assuming that a fraction of the total number of 
electrons was relativistic. The solutions from Paper~I that fit the observations 
of \object{Cyg~OB2~No.~9} could be divided into two groups: weak shock solutions 
with a constant relativistic electron fraction throughout the wind and strong 
shock solutions with a relativistic electron fraction increasing outward.

In the layered model there are no weak solutions (as can be seen in
Fig.~\ref{fig:solutions}a, $\chi>3.5$), which is an effect of
the cooling on the radio spectral index. To explain this, let us consider
the example of a multiple shock model given in Sect. \ref{sect:multipleshocks}.
This layered model fits the observations. It is also, to a good approximation, 
a continuous model in the sense that a large part of the wind is filled 
with relativistic electrons, with a fraction that is constant throughout the wind. 
Yet a true continuous model with the same parameters cannot fit the observations. 
The layered model has a steeper slope than the continuous model, because of 
the wavelength dependence of the width of the layer. Thus the weak shock solutions 
that were found in the continuous model cannot occur in the layered model.

For most solutions in the continuous model the fraction of relativistic
electrons increases further out in the wind. This means that the bulk of the 
synchrotron emission is produced at the outside of the synchrotron emitting 
region as it is in a single shock model. The extent of the synchrotron emitting 
region $R_{\rm max}$ in the continuous model is also comparable to the position 
of a single shock in the layered model. For $B_*=100$~G, $R_{\rm max}$ is 
between $520-750~R_*$. Therefore the continuous model and the layered model
agree on the location of the emission layer in the wind.

\section{Discussion and conclusions}\label{sect:conclusions}

In this paper we refined the model from Paper~I to take into account the 
cooling of the relativistic electrons and the large variety of shock strengths 
in the stellar wind. The cooling reduces the emitting volume to narrow layers 
behind the shock. Since the strongest shocks dominate the synchrotron 
emission produced, only a small number of layers are responsible for the 
observed flux. We further simplified the layered model by 
assuming that only one shock is present in the wind. Applied to a specific 
observation of \object{Cyg~OB2~No.~9}, the model produces meaningful constraints
on all parameters. The high values that we find for the 
shock velocity jump
(compared to hydrodynamical models) suggest emission by multiple shocks.
Note that there is an apparent contradiction between the large number of shocks 
needed to reduce the velocity jump and the small number of shocks expected 
to produce the bulk of the emission.

An important result of the single shock model is that the compression 
ratio ($\chi=3.5-4$) and the position of the shock ($R_S=520-640~R_*$) are 
well constrained. This means that strong shocks must survive up to large distances 
in the wind 
(a conclusion also reached by Chen \cite{C92}).
This is consistent with the results from Paper~I and from 
hydrodynamical calculations (Runacres~\&~Owocki~\cite{RO04}). The 
shock velocity jump
$\Delta u$ and the surface magnetic field
$B_*$ both have values which are not so well constrained and extend over a large 
range. However, these values are tightly related, so that the detection limit on 
$B_*$ gives a lower limit on $\Delta u\approx 290~{\rm km\,s^{-1}}$. 

The synchrotron emission is produced by the strongest shocks in the stellar
wind.  Because these shocks move through the wind, the radio spectrum
must change in time. This may explain the variability observed in non-thermal
emitters (Bieging et al.~\cite{BAC89}). In that case, the time-scales of 
the variability should be comparable to the flow time. When a single shock
travels from 450~$R_*$ to $550~R_*$, the synchrotron emission at 20~cm increases
from 0 to 4.4~mJy within a flow time of about 6 days.
Therefore the time-scale of variability in \object{Cyg~OB2~No.~9} should be of the order
of days to a week. However, 
it is unlikely that a shock would cover a complete solid angle, as it will probably 
be disrupted by lateral instabilities (e.g. Rayleigh-Taylor). This will reduce the 
amplitude of variability. Furthermore,
when the synchrotron emission is produced by multiple
shocks, the variations due to the changing positions of the shocks
could be
completely
averaged out.

The efficiency of inverse-Compton and adiabatic cooling
prevents electrons from carrying their energy very far from a shock. This however 
does not mean that the relativistic electrons need to be accelerated {\it in situ}.
For example, electrons accelerated near $300~R_*$ still emit a significant amount of
synchrotron radiation by the time they reach $600~R_*$ (see on Fig.~\ref{fig:emissionlayer} 
the emissivity at $597~R_*$ produced by electrons originally accelerated at 
$339~R_*$). Thus synchrotron emission is still produced far out in the wind 
even though shocks do not need to survive up to these distances.

The presence of relativistic electrons in the wind also leads to the generation
of non-thermal X-rays, due to the inverse-Compton scattering of photospheric UV photons
(e.g. Chen~\&~White~\cite{CW91}). Since there is an abundant supply of UV photons
near the stellar surface, the non-thermal X-ray emission is strongest in the 
inner wind where there is no observable radio emission.
The emission shows up as
a hard tail in the X-ray spectrum. Since the energy of the X-ray photons ($E$) is
related to the energy of the incident UV photons ($E_0$) as $E\approx \gamma^2 E_0$
(see Rybicki~\&~Lightman~\cite{RL79}), with $\gamma$ the Lorentz factor of the
scattering electron, electrons with momenta between $ 10-20$~MeV/c ($\gamma\approx 20-40$)
are responsible for the emission between $1-10$~keV. Since the non-thermal radio 
emission comes from electrons with momenta of order $\sim 100~{\rm MeV/c}$, the 
non-thermal X-ray emission comes from less
energetic electrons than the non-thermal radio emission. 
Even so, the non-thermal X-ray emission is dominated by the strongest shocks. In this regard
the conclusion by Rauw et al.~(\cite{R02}) that the compression ratio derived from the 
X-ray data is $\chi\approx 1.75$, is somewhat surprising.

In order to produce non-thermal radio emission, only a magnetic field and relativistic
electrons are required. Although a magnetic field and shocks to accelerate electrons
should both be present in any early-type star, not
all these stars are non-thermal emitters. A possible explanation is the {\em value} of
the magnetic field. In a low magnetic field less radiation is produced by the
relativistic electrons and the contribution of the synchrotron emission to the radio
flux would be negligible. This suggests that the thermal emitters have a lower surface
magnetic field than the non-thermal emitters (Bieging et al.~\cite{BAC89}).

The values found for $\Delta u$ in the single shock model are rather
high when compared to values derived from hydrodynamical models, especially 
because these high values are needed far out in the stellar wind.
If we re-interpret the emission as coming from multiple shocks, the 
shock velocity jump
does not have to be as high as for a single shock. 
For shocks in the outer wind, a 
shock velocity jump
of  $50~{\rm km\,s^{-1}}$ is
a reasonable value (e.g. Runacres \& Owocki \cite{RO04}). For $B_*=200~{\rm G}$ 
the observations are then produced by a reasonable number of shocks. In weaker fields, 
however, too many shocks are required (also, the Razin effect is stronger).
Our synchrotron models will be confronted with
hydrodynamical calculations in a subsequent paper.

Both the high $\Delta u$ value and the location of the shock at $600~R_*$
in a single shock model
can be interpreted as coming from a colliding wind binary. The synchrotron emission is then produced by 
relativistic electrons accelerated, not in wind-embedded shocks, but in a colliding-wind
shock  (Eichler \& Usov \cite{EU93}). Even though the present model is not appropriate
for colliding winds,
the position of the shock and the value of $\Delta u$ are close to those found 
in \object{WR~147} (Dougherty et al. \cite{DP03}). This means that, although there is no spectroscopic 
evidence for binarity, the observed non-thermal emission of \object{Cyg OB2 No. 9} is consistent with
a colliding-wind binary. For Wolf-Rayet stars, the relation between non-thermal 
radio emission and binarity is firmly established (Dougherty \& Williams \cite{DW00}). 
For O stars, the situation is less clear. About half of the non-thermal O stars are 
apparently single. However, recent spectroscopic studies show that one of 
the archetypical single non-thermal O stars, \object{Cyg OB2 No. 8A}, is in 
fact a binary (De Becker et al. \cite{DB04b}). Therefore, the possibility that 
\object{Cyg OB2 No. 9} is a binary cannot be excluded.

\begin{acknowledgements}
We thank S. Dougherty and J. Pittard for careful reading of the manuscript.
We thank the referee, R. White, for suggestions that improved the paper.
SVL gratefully acknowledges a doctoral research grant by the Belgian Federal  
Science Policy Office (Belspo). Part of this research was carried
out in the framework of the project IUAP P5/36 financed by Belspo.
\end{acknowledgements}

\end{document}